\documentclass[12pt]{article}
\usepackage[dvips]{epsfig}
\usepackage{cite}

\title{Fermion dispersion in axion medium}

\author{N.V.~Mikheev\thanks{E-mail address:mikheev@uniyar.ac.ru},
E.N.~Narynskaya\thanks{E-mail address:elenan@uniyar.ac.ru}\\
{\small\it Division of Theoretical Physics, Department of Physics,}\\
{\small\it Yaroslavl State University, Sovietskaya 14,}\\
{\small\it 150000 Yaroslavl, Russian Federation.}}

\date{}

\begin{document}

\maketitle

\begin{abstract}
The interaction of a fermion with the dense axion medium 
is investigated  for the purpose
of finding an axion medium effect on the fermion dispersion. 
It is shown that  axion medium  influence on the fermion 
dispersion under astrophysical conditions is negligible small if  the correct Lagrangian
of the axion-fermion interaction  is used. 

\end{abstract}

\def\beq{\begin{equation}}
\def\eeq{\end{equation}}
\def\bd{\begin{displaymath}}
\def\ed{\end{displaymath}}

\section{Introduction.}

\indent\indent 
The investigations of elementary particle physics beyond the standard model are one of the interest 
line of modern researches.  Particular attention  is given to the light goldstone and pseudogoldstone
bozons, arising at spontaneously  breaking of some symmetry. In view of weak  interaction
with matter the special interest for such particles is caused by possible astrophysical applications. In particular,
the nature of cool dark matter remains  the mystery of modern cosmology. The one of the most likely candidate for the dark matter is axion associated with Peccei-Quinn symmetry~\cite{Peccei,Wilczek}. The extremely small value of axion mass~\footnote{ We use natural units in which
$c=\hbar=1$.},

\beq 10^{-6}  \:\mbox{eV} < m_{a} < 10^{-2} \:\mbox{eV},
\label{eq:ax_mass}
\eeq
and gigantic life time in vacuum~\cite{Raffelt:90}
	\beq \tau_{a \to 2 \gamma} \sim  10^{50} \,\,
\mbox{sec} \,\,\left ( \frac{F_a}{10^{10} \, \mbox{GeV}} \right )^6 \,
\left ( \frac{\omega}{1 \, \mbox{MeV}} \right ),
\medskip \eeq
are two paramount factors why axion is of astrophysical and cosmology interest. Axion with such mass 
(\ref{eq:ax_mass}) could form a considerable fraction of dark matter. Moreover the region with high density of axion (so-called axion stars) could exist in Universe.The possibility existence of such  axion objects has been discussed~\cite{Hogan:88}. Dense axion matter playing the role of active medium  could influence on particles properties. In particular, the change of the dispersion relation of  the high energy cosmic ray propagating  in pseudoscalar particle medium was investigated in~\cite{Sahu}. However, as it will be shown further the result presented in~\cite{Sahu} is significantly overestimated. This is due to the fact that authors~\cite{Sahu} used the Lagrangian with the Yukawa coupling as the Lagrangian of axion-fermion interaction.

In this letter we review the influence of cool axion matter on the  fermion dispersion. The modification of the
 fermion dispersion is caused by fermion "forward-scattering" of medium axion.  The axion-fermion interaction 
 will be analyzed by using the Lagrangian in the form~\cite{Raffelt:96}
\medskip
\begin{equation}
	L_{af} = \frac{c_a}{F_a}\:\left(\overline{\psi}\,\gamma_\mu \,
\gamma_5 
\,\psi \right)\;\partial_\mu a,
	\label{eq:lag-1}\medskip
\end{equation}
where  $c_a$ is the model-dependent parameter of the order of unit,
$F_a$ is the Peccei-Quinn symmetry-breaking scale,
$\psi$ and $a$ are the fermion and axion fields respectively. It should be noted that
 in order to obtain the  
correct amplitudes of processes involving two axions,
 it is necessary to use just the Lagrangian with  derivative (\ref{eq:lag-1}) 
rather then  Yukawa Lagrangian
\medskip
  \beq L_{af} = \frac{- 2 i c_a m_f}{F_a}\:\left(\overline{\psi}\!\,\gamma_5 \!\,\psi  \right)\; a,
	\label{eq:lag-2} \medskip
	\eeq
as it was indicated for the first time in~\cite{Raffelt:88}.

 \section{Dispersion of fermion in axion medium.}

\indent\indent  
For investigation the fermion dispersion in axion medium the process of  
fermion "forward  scattering" of plasma axions (fig.1) should be taken into account. 
The amplitude of this process is intimately connected with fermion mass operator 
$\sum$ by the following relation
\medskip
\beq
 M = - \bar u(p) \sum u(p),
 \label{eq:mass}\medskip
\eeq
 where $u(p)$ is the bispinor amplitude of fermion.
\begin{figure}[b]
\medskip
\medskip
\centerline{\includegraphics{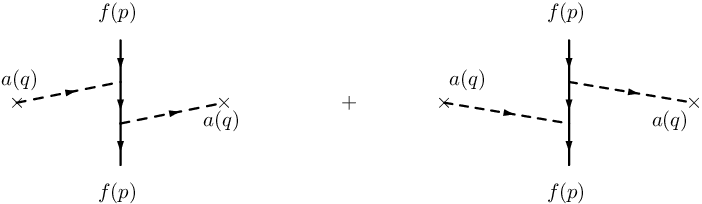}}
\medskip
\medskip
\caption{The diagrams
describing   fermion 
  "forward  scattering" of  plasma axions.}
\label{fig:fig2}
\end{figure}

The amplitude corresponding to the left diagram in fig.1
has the form
\medskip
\begin{equation}
M^{(1)} = \frac{i c_a^2}{2 F_a^2 }\!\int\frac{d^3 q}{(2\pi)^3 \, \omega}
 \:f(\omega)\:
\left ( \bar u(p) \:\left[\hat q \gamma_5 S(p+q) \hat q \gamma_5 \right]\:u(p)\right),
\label{eq:amp-1}
\medskip \end{equation}
where $q^\mu = (\omega, \vec q)$ and $p^\mu = (p_0, \vec p)$ are the axion and fermion four-momentum 
correspondingly, $\hat q = (q_\mu \gamma^\mu)$,
$f(\omega)$ is the axion distribution function, $f(\omega) = (e^{\omega/T}-1)^{-1}$,
$S(p)$ is the  propagator of fermion with  mass $m_f$:
$$ S(p) = \frac{i (\hat p + m_f)}{p^2 - m_f^2}. 
$$

The amplitude corresponding to the right diagram in fig.1 can be obtained from (\ref{eq:amp-1}) by the substituting $q \to - q$.  Performing  some calculations,  the total amplitude of the coherent scattering process $f + a \to f + a$ 
  of all plasma axions can be written as
\medskip
\begin{equation}
M =  \frac{  c_a^2 }{2 F^2_a }\int\frac{d^3q}{(2\pi)^3 \, \omega}\: f(\omega) 
\:\bar u(p) \left( \frac{(p^2-m_f^2)\hat q + (\hat p + m_f)m_a^2}{(p+q)^2 - m_f^2} \;+\;q \to -q \right)u(p).
\label{eq:amp-2}
\medskip 
\end{equation}

With the definition (\ref{eq:mass})  the fermion mass operator has the form 
\medskip
\begin{equation}
\sum  =  \frac{- c_a^2 }{2\!F^2_a }\int\frac{d^3q}{(2\pi)^3\, \omega}
\: f(\omega) 
\:\left( \frac{(p^2-m_f^2)\hat q + (\hat p + m_f)m_a^2}{(p+q)^2 - m_f^2} \;+
\;q \to -q \right).
\label{eq:s-1}
\medskip \end{equation} 

There is good reason to believe that axion gas is most likely a cool nonrelativistic medium. So it is possible
to carry out  calculations in the approximation   $q^\mu \simeq (m_a, \vec q), \,\,\mid \vec q\mid << m_a$. 
Under such conditions the expression for the mass operator is essentially simplified

\medskip
\begin{equation}
\sum  =  \frac{- c_a^2 n_a}{4 F^2_a p_0^2 m_a}\:  \left\{2 p_0 \hat U  (p^{2} - m_f^2) 
- (\hat p + m_f)(p^{2} - m_f^2 + m_a^2)  \right\},
\label{eq:s-2}
\medskip \end{equation}  
where $n_a$ is the concentration of axion gas,
 $U^\mu$ is the four-vector of medium which in its rest system 
 has the form $U^\mu = (1,0,0,0)$.

As one  see from (\ref{eq:s-2}) the mass operator can be presented in the form: 
\medskip
\begin{equation}
\sum  =  A m_f + B \hat p + C \hat U,
\label{eq:s-3}\medskip
\end{equation}
\medskip
$$ A = B = \frac{n_a c^2_a(p^{2} - m_f^2 + m_a^2)}{4F_a^2 p_0^2 m_a}, 
\qquad C = \frac{- n_a c^2_a(p^{2} - m_f^2)}{2 F_a^2 p_0 m_a}.$$
\medskip

The fermion dispersion low can be extracted from equation
\medskip
\beq
det \left( \hat p - m_f - \sum  \right) = 0.
\label{eq:disp}
\eeq
\medskip

For the fermion energy in the cool axion medium  from (\ref{eq:disp}) with (\ref{eq:s-3})
one obtains
\medskip
\begin{equation}
E_{f} \simeq   \sqrt{\vec p^{\,\,2} + m_f^2} + \frac{m_f^2(A+B)}{\sqrt{\vec p^{\,\,2} + m_f^2}} - C,
\label{eq:energy}
\medskip
\end{equation}
where $\sqrt{{\vec p}^{\,\,2} + m_f^2}=E_0$ is the fermion energy in vacuum.

This is our main result  determining the fermion dispersion in the cool axion medium. 
For the purpose of comparison (\ref{eq:energy}) with the previous result \cite{Sahu} we estimate the deviation of fermion mass squared in the cool axion medium from vacuum one, $\delta m_f^2$:

$$ 
\delta m_f^2 = p^2 - m_f^2 = E_f^2 - E_0^2 = E_f^2 - {\vec p}^{\,\,2} - m_f^2.$$

The deviation of  fermion mass  squared from (\ref{eq:energy}) can be written as
\beq 
\delta m_f^2 \simeq 2 m_f^2(A+B) - 2 C \sqrt{\vec p^{\,\,2} + m_f^2}.
\eeq

Taking into account smallness of the parameters $A,B$ and $C$ 
the axion medium correction to squared of the fermion mass is  
\medskip
\begin{equation}
\delta m_f^2 \simeq \frac{n_a m_a c_a^2 m_f^2}{({\vec p}^{\,\,2} + m^2_f) F_a^2}
\label{eq:m-1}.
\end{equation} 

It can be seen that this contribution is strongly suppressed not only by the energy scale of the breakdown
 symmetry $F_a$ in denominator but the axion mass $m_a$ in numerator as well. 
As was pointed above, the result for the squared of 
the fermion mass in axion medium under the same physical conditions 
was presented early~\cite{Sahu} in the form
\medskip
\begin{equation}
(\delta m_f^{2})^{Sahu} \simeq \frac{n_a m_f^2}{2 m_a F_a^2}. 
\label{eq:m-pap} \medskip
\end{equation}

The comparison of  (\ref{eq:m-1}) with  (\ref{eq:m-pap}) 
\medskip
$$ R = \frac{\delta m_f^2}{(\delta m_f^2)^{Sahu}}  = 
\frac{2 c_a^2 m_a^2}{E_0^2} << 1$$
\medskip
illustrated that the result of the paper~\cite{Sahu} is strongly overestimated. This is due to the fact
authors \cite{Sahu} uncorrectly used   
the Lagrangian with pseudoscalar Yukawa coupling for  investigation the process of fermion "forward scattering" of axions medium.
\section{Conclusion}
\indent\indent 
 We have studied the influence of cool axion medium to the fermion dispersion. The contribution into squared of fermion mass due to presence of axion medium is calculated. The investigation of 
 this type were performed early in paper~\cite{Sahu}. In this paper we demonstrated that  calculation of fermion
 dispersion in the axion medium with Yukava Lagrangian leads to the  significantly overestimate results. Under  real astrophysical conditions the influence of axion medium to the fermion dispersion turns out to be 
 negligible small.

\vspace{5mm}
This work supported in part by the Council on Grants by the President of
 Russian Federation for the Support of Young Russian Scientists and Leading
 Scientific Schools of Russian Federation under the Grant No.
 NSh-497.2008.2 and by the Russian Foundation for Basic Research 
under the Grant No. 07-02-00285-a.
\vspace{5mm}


\end{document}